\documentclass[%
  aip,
  amsmath,amssymb,
  reprint,%
 ]{revtex4-2}

    \usepackage[T1]{fontenc}
\usepackage[utf8]{inputenc}
\usepackage[english]{babel}
\usepackage[]{graphicx}
\usepackage{amsmath}
\usepackage{amssymb}
\usepackage{amsfonts}
\usepackage[]{units}
\usepackage{times}
\usepackage{physics}
\usepackage{bm}
\usepackage{bbm}
\usepackage{dcolumn}
\usepackage{soul}		
\usepackage{xcolor}
\usepackage{tabularx}
\usepackage{hyperref}
 \hypersetup{
     colorlinks=true,
     linkcolor=blue,
     filecolor=blue,
     citecolor = blue,      
     urlcolor=blue,
     }

\newcommand{\ch}[1]{{#1}} 
\newcommand{\comment}[1]{}

\def\pM{\ensuremath{%
  \genfrac{}{}{0pt}{}{+}%
    {\scriptstyle(\kern-1pt-\kern-1pt)}}}

\renewcommand{\emph}{\textit}

\usepackage{etoolbox}

\makeatletter
\def\@email#1#2{%
 \endgroup
 \patchcmd{\titleblock@produce}
  {\frontmatter@RRAPformat}
  {\frontmatter@RRAPformat{\produce@RRAP{*#1\href{mailto:#2}{#2}}}\frontmatter@RRAPformat}
  {}{}
}%
\makeatother

\begin{document}
\preprint{AIP/123-QED}

\title{Theory of radial oscillations in metal nanoparticles driven by optically induced electron density gradients}
\author{Robert Salzwedel}
\email{r.salzwedel@tu-berlin.de}
\affiliation{Institut für Theoretische Physik, Nichtlineare Optik und Quantenelektronik, Technische Universität Berlin, Berlin, Germany}
\author{Andreas Knorr}
\affiliation{Institut für Theoretische Physik, Nichtlineare Optik und Quantenelektronik, Technische Universität Berlin, Berlin, Germany}
\author{Dominik Hoeing}
\affiliation{Institut für Physikalische Chemie, Universität Hamburg, Hamburg, Germany}
\affiliation{The Hamburg Centre for Ultrafast Imaging, Hamburg, Germany}
\author{Holger Lange}
\affiliation{Institut für Physikalische Chemie, Universität Hamburg, Hamburg, Germany}
\affiliation{The Hamburg Centre for Ultrafast Imaging, Hamburg, Germany}
\author{Malte Selig}
\affiliation{Institut für Theoretische Physik, Nichtlineare Optik und Quantenelektronik, Technische Universität Berlin, Berlin, Germany}

\date{\today}

\begin{abstract}
We provide a microscopic approach to \ch{describe} the onset of radial oscillation of a \ch{silver} nanoparticle. Using the Heisenberg equation of motion framework, we find that the coupled ultrafast dynamics of coherently excited electron occupation and the coherent phonon amplitude \ch{initiate} periodic size oscillations of the nanoparticle. Compared to the established interpretation of experiments, our results show a more direct coupling mechanism between the field intensity and \ch{coherent} phonons. This interaction triggers a size oscillation \ch{via an optically induced electron density gradient occurring} directly with the optical excitation\ch{. This source} is more \ch{efficient} than the incoherent heating process currently discussed in the literature and well-describes the early onset of the oscillations in \ch{recent} experiments.
\end{abstract}
\maketitle

\section{Introduction}
\begin{figure}[h!]
  \includegraphics[width=\linewidth]{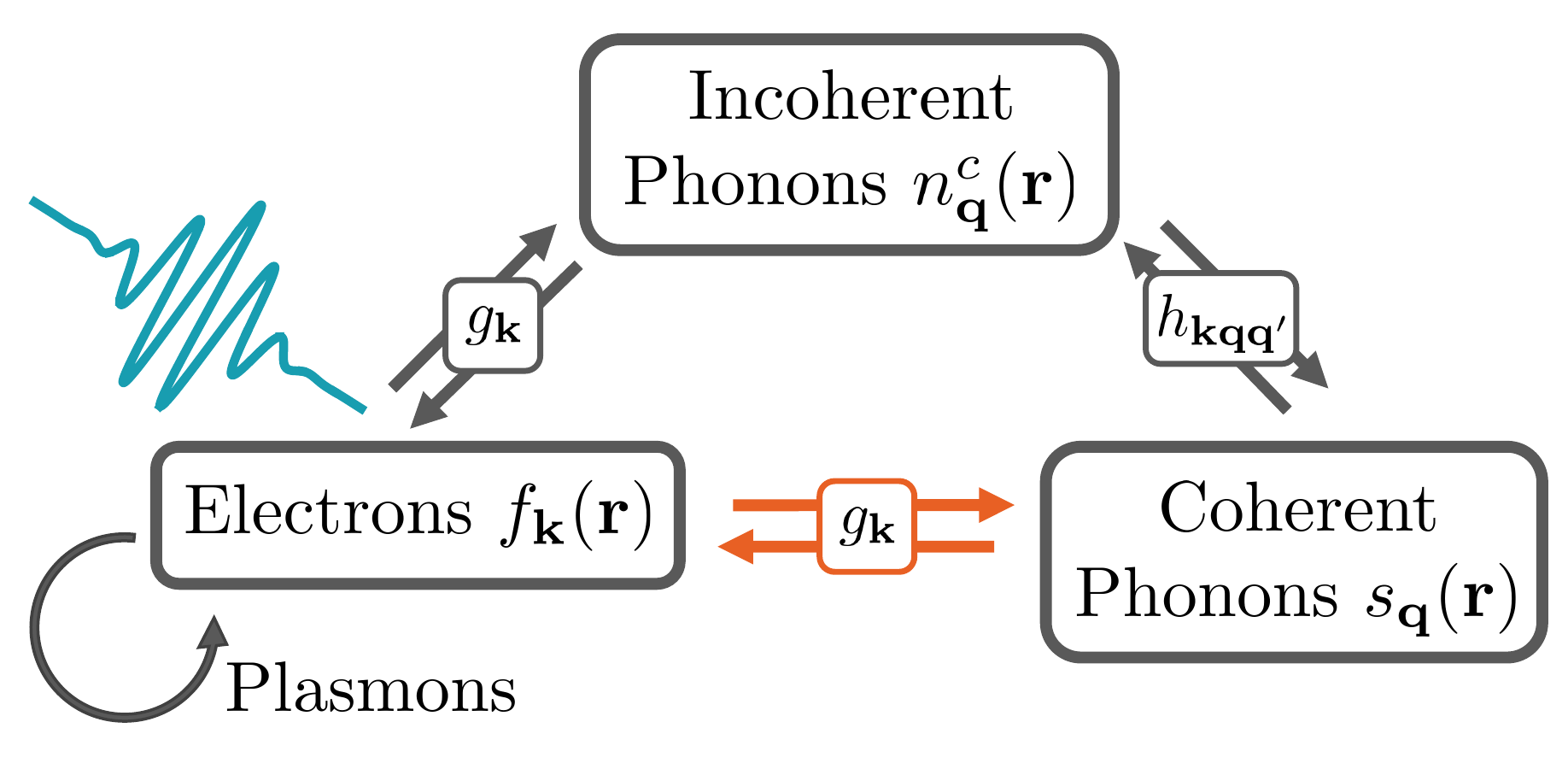}
  \caption{\textbf{Interaction scheme.} Incident light excites the electrons that self-interact via a collective plasmon response. In all classical theories (gray arrows), the electrons couple to incoherent phonons, i.e.,~temperature, which transiently drive the coherent oscillations due to the spatial restriction (spherical particles). We study the possibility of a direct interaction of electrons and coherent phonons (orange arrows).}
  \label{fig:interaction}
\end{figure}

In the late 19th century, a special focus of theoretical physics was the study of vibrations in elastic spheres. 
Sir Horace Lamb  provided the first fundamental solutions showing that tangential modes occur in addition to simple breathing modes.\cite{lamb1881vibrations} 
A century later, such breathing modes were observed in metal nanoparticles (MNPs) following the optical excitation of localized surface plasmons.\cite{hartland2002coherent,hodak1999size,ng2011laser,brongersma2015plasmon,crut2015acoustic,vandijk2005detection} 
A detailed study of the occurring MNP-size oscillations confirmed Lamb's calculations, as the measured oscillation frequency matched perfectly the theoretically predicted one.\cite{crut2015acoustic} 
However, few studies have investigated the actual driving source of these size oscillations because their onset is difficult to access experimentally. 
It has long been assumed that the optically heated electron gas transfers its energy to the phonons in the nanoparticle, resulting in rapid expansion. 
This thermal effect was expected to be the dominant driving source of the size oscillations.\cite{hartland2002coherent,boriskina2017losses,besteiro2019fast}\\
However, recent advances in experimental techniques have allowed a more detailed study of the temporal onset of nanoparticle oscillations\cite{kim2019ultrafast,clark2015imaging} and seem to support the idea of a further driving source in addition to the thermal expansion. 
In particular, there is evidence that the radial oscillations begin before the expected timescales of lattice heating. 
Therefore, it occurs to be possible that the sequence of events described above (absorption, heating, and oscillation) does not prevail in the early stages after excitation and that the actual driving source has not yet been identified.\\
In the classical model, the current understanding\cite{hodak2000photophysics,hartland2002coherent} does not allow for a direct interaction of electrons and coherent phonons. 
Fig.~\ref{fig:interaction} illustrates that the interaction in the classical model (gray arrows) is always mediated by incoherent (thermal) phonons, which are described by temperature and not by coherent phonon oscillations. 
In order to match the experimentally observed oscillation phase, some publications\cite{hartland2002coherent,crut2015acoustic} add a hot electron pressure term via the electronic Grüneisen parameter, which allows for a more direct interaction of the hot electron system with the vibrational mode, initiating the oscillation onset earlier than the interaction mediated purely via thermal phonons.\\
In this work, we present a new quantum mechanical coupling mechanism that allows us to distinguish between coherent and incoherent phonon modes.
The ansatz is inspired by research from the semiconductor community.\cite{kuznetsov1994theory,scholz1993density,rossi2002theory,zeiger1992theory,wigger2014energy,kabuss2012optically,karwat2018coherent,krauss1997coherent,czerniuk2017picosecond} 
While the coherent phonon modes can be associated with radial vibrations of the macroscopic metal nanoparticle, the incoherent modes can only contribute on the timescale of thermalization of the electron-phonon system. 
Since our description is based on a Hamiltonian formulation, it allows for a consistent description of carrier-phonon coupling between all phonon modes and electrons. 
In particular, a direct coupling of optically induced electron density gradients and coherent phonon oscillations (orange arrows in Fig.~\ref{fig:interaction}) is possible, which is expected to lead to an oscillation onset directly with the optical excitation.\\

In this contribution, to study spatially resolved direct electron-phonon coupling in a nanoparticle, we develop a quantum theory that describes incoherent and coherent electron and phonon dynamics in nanoparticles after optical excitation. 
We employ the Heisenberg equation of motion framework to derive microscopic equations for the Wigner distributions of electrons and phonons. 
To reduce complexity, we explore the hydrodynamic limit of these equations by a coarse-graining on macroscopic observables. 
In doing so, we retain coherent variables such as the optical field and coherent phonons that interact with the macroscopic motion of charge density, current density, and temperature dynamics. 
By projecting our equations of motion onto the fundamental Lamb modes\cite{lamb1881vibrations,love1892treatise} of the elastic sphere under consideration, we can explicitly calculate the electron and phonon dynamics after optical pumping.\\
A new driving term, namely the optically induced spatial gradient of the electron density, is identified as the actual source of the radial breathing oscillations, which is present as long as the pump pulse is present in the system. Since the theory developed is restricted to a pure single-band model, silver is chosen as the material system because the intraband and interband resonances are spectrally well separated\cite{bonafe2017plasmon} in this material system and can be described with a pure intraband theory.\\
Our manuscript is organized as follows: We begin by introducing the microscopic model for the interaction of electron and phonon modes including electron-light coupling in Sec.~\ref{sec:microscopic}. 
To simplify the treatment, the dynamics are coarse-grained, leading to a closed set of coupled hydrodynamic equations in Sec.~\ref{sec:factorization}. 
In Sec.~\ref{sec:linearization}, the resulting nonlinear equations are linearized up to second order in the excitation field, which allows us to project onto the solutions of the vibrational Lamb modes in Sec.~\ref{sec:lamb}. 
In Sec.~\ref{sec:numerics}, we provide a numerical solution for the coupled ordinary differential equations that arise. 
Finally, we conclude with a comparison of the influence of the driving terms on the nanoparticle oscillation in Sec.~\ref{sec:conclusion}.

\section{Microscopic Approach}
\label{sec:microscopic}
%
The starting point for our investigation is the Hamiltonian 
\begin{align}
    H 
   & = 
    \sum_{\vb{k}}
        \varepsilon_{\vb{k}}
        a_{\vb{k}}^\dagger
        a_{\vb{k}}
    +\sum_{\vb{q}}
        \hbar \omega_{\vb{q}}
        b_{\vb{q}}^\dagger 
        b_{\vb{q}}
    \label{eq:hamiltonian}\\
    & + ie\sum_{\vb{k},\vb{K}}\mathbf{E}_{-\vb{K}}\cdot\nabla_{\vb{K}}(a_{\vb{k}}^\dagger a_{\vb{k}+\vb{K}})\nonumber \\
    & + 
    \sum_{\vb{q}\vb{k}}
        g_{\vb{q}} 
        a_{\vb{k}+\vb{q}}^{\dagger}
        a_{\vb{k}}
        \qty(
            b_{\vb{q}}
            +b_{-\vb{q}}^{\dagger}
            )
    + \frac 12 \sum_{\vb{k,k',q}}V_{\vb{q}}a^\dagger_{\vb{k+q}}a^\dagger_{\vb{k'-q}}a_{\vb{k'}}a_{\vb{k}}
   \nonumber  \\
     & + 
    \sum_{\vb{q}_1\vb{q}_2\vb{q}_3}
        h_{\vb{q}_1\vb{q}_2\vb{q}_3} 
        \qty(
            b_{\vb{q}_1}
            +b_{-\vb{q}_1}^{\dagger}
            )\qty(
            b_{\vb{q}_2}
            +b_{-\vb{q}_2}^{\dagger}
            )\qty(
            b_{\vb{q}_3}
            +b_{-\vb{q}_3}^{\dagger}
            ).\nonumber
\end{align}
The first term accounts for the dispersion of metal nanoparticle electrons $\varepsilon_\mathbf{k}$ with electronic annihilation (creation) operators $a^{(\dagger)}_\mathbf{k}$ with momentum $\mathbf{k}$ in a single band approximation, which suffices for intraband transitions.\cite{christensen1972band} 
For simplicity, we assume a parabolic dispersion $\varepsilon_\mathbf{k}=\frac{\hbar^2 \mathbf{k}^2}{2 m }$ with the effective electron mass $m$.\cite{johnson1972optical,yang2015optical} 
The second term incorporates the dispersion of dominant longitudinal acoustic (LA) phonons with phonon annihilation (creation) operators $b_\mathbf{q}^{(\dagger)}$ of momentum $\mathbf{q}$.  
In Debye approximation,\cite{madelung1978introduction,czycholl2017theoretische} the appearing phonon dispersion is assumed to be linear $\omega_\mathbf{q} = c_{LA} |\mathbf{q}|$ with the velocity of sound $c_{LA}$.\cite{drexel1969phonon} 
The second line describes the intraband part of the semiclassical light-matter coupling\cite{hannes2019higher,haug2009quantum} with the Fourier component $\mathbf{E}_\mathbf{K}(t)$ of the exciting electric field $\mathbf{E} (\mathbf{r},t) = \sum_\mathbf{K} \exp{i \mathbf{K}\cdot \mathbf{r}}\mathbf{E}_\mathbf{K}(t)$ with the elementary charge $e$. 
The first term in the third line considers the electron-phonon interaction with the electron-phonon coupling strength 
$g_\mathbf{q}=
	\sqrt{\flatfrac{\hbar N}{2M\omega_{\vb{q}}} }
    i q V_{0} 
$,\cite{czycholl2017theoretische}
where $M$ is the ion mass in the unit cell and $N$ is the ion number in the crystal.
$V_0$ is the strongly screened version of the Fourier transformed electron-ion potential.\cite{kittel1987quantum} 
The second term in that line accounts for the carrier-carrier interaction via Coulomb interaction $V_q = e^2/\varepsilon_0 \Omega \,q^2$ of the individual carriers with momentum exchange $\vb{q}$ between two carriers with momenta $\vb{k}$ and $\vb{k'}$ and the crystal volume $\Omega$. 
The fourth line describes phonon-phonon interaction that arises from anharmonic corrections of the phonon mode expansion. 
The appearing matrix element conserves the total momentum of the involved phonons, i.e., $h_{\vb{q}_1\vb{q}_2\vb{q}_3} = \Tilde{h}_{\vb{q}_1\vb{q}_2\vb{q}_3} \delta_{\vb{q}_1+\vb{q}_2+\vb{q}_3,0}$. 
All parameters used for the numerical evaluation, including the material-specific silver parameters, are listed in Tab.~\ref{table_parameters} in Sec.~\ref{sec:numerics}.\\ 
To investigate the origin of the coherent oscillations in the silver nanoparticles, we study the lattice displacement $\vb{u} (\mathbf{R}_n,t)$ at the lattice vector $\vb{R}_n$ of the MNP.\cite{haken2013quantenfeldtheorie,czycholl2017theoretische,madelung1978introduction,kuznetsov1994theory}
It can be expressed in terms of phonon annihilation (creation) operators $b_{\vb{q}}^{(\dagger)}$ as:\cite{mahan2013many,gross1986many,madelung1978introduction,wigger2014energy}
\begin{align}
      \vb{u} (\mathbf{R}_n,t)
    &=
        \sum_{\vb{q}}
            \sqrt{ \frac{\hbar}{2 M N \omega_\mathbf{q} } }
            \mathbf{e}_\mathbf{q} 
            e^{ i \vb{q} \cdot \vb{R}_n}
            \qty(\expval{b_{\vb{q}}}+\expval*{b_{-\vb{q}}^{\dagger}}).
            \label{eq:displacement}
\end{align}
This expression describes the lattice displacement for vibrational modes in the MNP, in particular, the breathing modes, which can be detected in optical experiments, i.e., x-ray scattering or transient absorption spectroscopy, as a direct observable. 
$\mathbf{e}_\mathbf{q}$ is the phonon polarization vector of $LA$ phonons.\cite{gross1986many} 
Inspired by the definition of the macroscopic lattice displacement in Eq.~\eqref{eq:displacement}, we define the coherent phonon amplitude
$
    s_{\vb{q}}\equiv\frac 12\qty(\expval{b_{\vb{q}}}+\expval*{b_{-\vb{q}}^{\dagger}}).
$
\cite{kuznetsov1994theory,scholz1993density,rossi2002theory,sanders2013theory}
We obtain an equation of motion for the phonon annihilation (creation) operators $b_{\pm\vb{q}}^{(\dagger)}$ by exploiting the Heisenberg equation of motion with the Hamiltonian [Eq.~\ref{eq:hamiltonian}],
\begin{align}
	&\left( i \hbar \partial_t \mp \hbar \omega_{\pm\mathbf{q}}  \right) b_{\pm\vb{q}}^{(\dagger)}
	=\pm
	\sum_{\vb{k}}
	g_{-\vb{q}}
	a^\dagger_\mathbf{k-q} a_\mathbf{k}\label{eq:eomcopho1}\\ 
	&\pm 3\sum_{\vb{q'}} \Tilde{h}_{\vb{q,q',q-q'}} \qty[b_{\vb{q'}}+b^\dagger_{-\vb{q'}}]\qty[b_{\vb{q-q'}}+b^\dagger_{-\vb{q}+\vb{q'}}]\nonumber,
\end{align}
where the lower line corresponds to the creation and the upper to annihilation operators.
The left-hand side includes the dispersion of the phonons $\omega_{\vb{q}}$. 
The right-hand side contains the sources of the phonon amplitudes. 
The first term represents a momentum transfer from the electronic system to the phonons. 
The second term originates from phonon-phonon interaction. 
In the next step, we apply a time derivative to Eq.~\eqref{eq:eomcopho1}, which results in a second-order differential equation for the coherent phonon amplitude,
\begin{widetext}
\begin{align}
\label{eq:eom_coherent_phonon}
    \qty(\partial_t^2+\omega_{\vb{q}}^2)s_{\vb{q}}= 
	&-\frac{\omega_{\vb{q}}g_{-\vb{q}}}{\hbar}\sum_{\vb{k}} \Tilde{f}_{\vb{k}}(\vb{q})
	-\frac{3\omega_{\vb{q}}}{\hbar}\sum_{\vb{q'}}
	\Tilde{h}_{\vb{q,q',q-q'}}\qty[2\Tilde{n}_\mathbf{q'}^c (\mathbf{q})+\delta_{\vb{q},0}
	]\\\nonumber
	&-\frac{3\omega_{\vb{q}}}{\hbar}\sum_{\vb{q'}}\Tilde{h}_{\vb{q,q',q-q'}}
	\qty[
	    4 s_{\vb{q'}}s_{\vb{q-q'}}+
	    \expval{b_{\vb{q'}}b_{\vb{q-q'}}}^c
	    +\expval{b_{\vb{-q'}}^\dagger b_{\vb{-q+q'}}^\dagger}^c
	].
\end{align}
\end{widetext}
On the right-hand side, we can identify two sources for the coherent phonon amplitude. 
The first term is the Fourier transform of the Wigner function of the electrons $\Tilde{f}_\mathbf{k} (\mathbf{q}) = \expval*{a^\dagger_\mathbf{k-q} a_\mathbf{k}}$.\cite{hess1996maxwell} 
The momentum- and real-space-dependent Wigner distribution is accordingly given by $f_\mathbf{k} (\mathbf{r}) = \sum_\mathbf{q} e^{i \mathbf{q}\cdot \mathbf{r}} \Tilde{f}_\mathbf{k} (\mathbf{q})$. 
The second term considers driving of the coherent phonon amplitude from the incoherent phonon distribution $\Tilde{n}_\mathbf{q'}^c (\mathbf{q}) \equiv \expval*{b^\dagger_\mathbf{q'-q} b_\mathbf{q'}}-\expval*{b_{\vb{q'-q}}^\dagger}\expval{b_{\vb{q'}}}$.\cite{rossi2002theory,kuznetsov1994theory} 
The last term in the first line is a Kronecker symbol that results from the commutation of the phonon creation (annihilation) operators to the normal order. 
It resembles a constant force on the oscillator that results in a shift of the oscillator's equilibrium position due to phonon-phonon interaction. 
In the second line, the first term describes a self-interaction and scales quadratically with the coherent phonon amplitude $s_{\vb{q}}$.
Since the equations of motion are not closed, we derive an equation of motion for the last two terms in the last line. 
After a Markov approximation and a second-order Born decoupling, we find that they couple to phonon creation (annihilation) operators $b_{\vb{q}}^{(\dagger)}$ so that the phonon coherences occur as dephasing and can be written as
\begin{align}
    \qty(\partial_t^2+2\gamma_{\vb{q}} \partial_t+\omega_{\vb{q}}^2)s_{\vb{q}}&= 
	-\frac{\omega_{\vb{q}}g_{-\vb{q}}}{\hbar}\sum_{\vb{k}} \Tilde{f}_{\vb{k}}(\vb{q})\label{eq:microscopic_oscillator}\\\nonumber
	&-\frac{3\omega_{\vb{q}}}{\hbar}\sum_{\vb{q'}}
	\Tilde{h}_{\vb{q,q',q-q'}}\qty[2\Tilde{n}_\mathbf{q'} (\mathbf{q})+\delta_{\vb{q},0}
	].
\end{align}
This equation is formally equivalent to a classical damped oscillator equation for mode $\vb{q}$. 
The left-hand side describes the oscillation of the coherent phonon amplitude with damping rate $\gamma_{\vb{q}}$, resulting from phonon-phonon interaction, which is given in Eq.~\eqref{eq:phonon_damping}, and the oscillator frequency $\omega_{\vb{q}}$. 
In Eq.~\eqref{eq:microscopic_oscillator} and in the following, non-linear terms in the coherent phonon amplitude $s_{\vb{q}}$ are suppressed since weak excitation is assumed.
As the electron Wigner function $\Tilde{f}_{\vb{k}}(\vb{q})$ acts as a driving term in the coherent phonon oscillation [Eq.~\eqref{eq:microscopic_oscillator}], we also derive an equation of motion for the electronic Wigner function within the gradient expansion\cite{hess1996maxwell,breusing2011ultrafast} and obtain
\begin{widetext}
    \begin{align}
    &\partial_t f_{\vb{k}}(\vb{r},t) +\nabla_{\vb{r}}f_{\vb{k}}(\vb{r},t)\cdot \qty[\vb{v}_{\vb{k}}
    +\frac{e}{\hbar}\nabla_{\vb{k}}\Phi_{\vb{k}}^F(\vb{r})] 
    = \label{eq:eomwignere} 
    \nabla_{\vb{k}}f_{\vb{k}}(\vb{r},t) \cdot 
    \qty[
        \frac{e}{\hbar} 
        \qty(
            \vb{E}(\vb{r},t)
            -\partial_{\vb{r}}\Phi^H(\vb{r})
            +\partial_{\vb{r}}\Phi_{\vb{k}}^F(\vb{r})
            ) 
        +\frac i \hbar \sum_{\vb{q}} g_{\vb{q}}e^{i \vb{q}\cdot\vb{r}}\vb{q}\,s_{\vb{q}}(t)
        ]\\
    &+\sum_\mathbf{q} 
    \qty[
        \Gamma^{\text{in}}_\mathbf{k+q,k}(\mathbf{r},t) (1-f_\mathbf{k}(\mathbf{r},t)) 
        - \Gamma^{\text{out}}_\mathbf{k,k+q}(\mathbf{r},t) f_\mathbf{k}(\mathbf{r},t)
        ]
    +\sum_\mathbf{q} 
    \qty[
        \mathcal{W}^{\text{in}}_\mathbf{k+q,k}(\mathbf{r},t) \qty(1-f_\mathbf{k}(\mathbf{r},t)) 
        - \mathcal{W}^{\text{out}}_\mathbf{k,k+q}(\mathbf{r},t) f_\mathbf{k}(\mathbf{r},t)
    ].\nonumber
\end{align}
\end{widetext}
 The left-hand side considers the drift of the electronic Wigner function with group velocity $\mathbf{v}_\mathbf{k} = \nabla_\mathbf{k} \varepsilon_\mathbf{k}$ under the effect of the Fock potential $\phi_{\vb{k}}^F(\vb{r})$ that is given in Eq.~\eqref{eq:fock_potential}. 
 The first term on the right-hand side accounts for the acceleration of electrons in the external optical driving field $\vb{E}(\vb{r},t)$ under the additional Hartree $\Phi^H(\vb{r})$ and Fock potentials $\Phi_{\vb{k}}^F(\vb{r})$. 
 They result from the inclusion of the Coulomb contribution in the Hamiltonian and are derived in a mean-field and gradient approximation. 
 In App.~\ref{sec:hartree_fock}, we explain in more detail how the Hartree-Fock contributions enter into the equation and provide definitions of the Hartree and Fock potential in Eq.~\eqref{eq:hartree_potential} and Eq.~\eqref{eq:fock_potential}, respectively. 
 The third term on the right-hand side describes the acceleration of electrons due to interaction with coherent phonons $s_{\vb{q}}(\vb{r},t)$. 
 The Fock contributions are included here for completeness reasons but will be neglected in the following as quantum corrections to a semiclassical hydrodynamic model as they are minor corrections to the total field $\Tilde{\vb{E}} \equiv \vb{E}-\partial_{\vb{r}}\Phi^H$ combining the external field $\vb{E}$ and the Coulomb Hartree corrections of the internal field $-\partial_{\vb{r}}\Phi^H$. 
 The first term in the second line on the right-hand side was derived using a second-order Born-Markov approximation for the carrier and phonon dynamics and takes into account the electron-phonon kinetic scattering in the limit of the standard Boltzmann equation.\cite{hess1996maxwell,jago2019spatio,selig2020suppression,venanzi2021terahertz} 
 The last term on the right-hand side accounts for electron-electron scattering events and can be derived in terms of a second-order correlation expansion.\cite{hess1996maxwell,lindberg1988effective,binder1992carrier,selig2020suppression,venanzi2021terahertz} 
 The scattering terms depend on the occupation of the initial state of the scattering event, i.e., in the case of in-scattering, on the electron occupation terms $f_{\vb{k}}$, and in the case of out-scattering, on the hole occupation terms $(1-f_{\vb{k}})$. 
 The scattering rates are provided in App.~\ref{sec:scatteringrates}.
 
\section{Hydrodynamic approach}
\label{sec:factorization}
%
In order to simplify the treatment of the coupled electron-phonon dynamics in the MNP, we perform a momentum expansion of the Wigner function \cite{grad1949kinetic,gardner1994quantum,cai2012quantum} in the group velocity $\vb{v}_{\vb{k}}$. 
This way, the microscopic information described by the occupations at any wavevector of electrons $\mathbf{k}$ and phonons $\mathbf{q}$ is coarse-grained.
The momenta of the Wigner function can be identified with the macroscopic observables charge density $\rho(\vb{r},t)$ and current density $\vb{j}(\vb{r},t)$,\cite{cai2012quantum,grad1949kinetic,gardner1994quantum}
\begin{align}
    \rho (\mathbf{r},t) &\equiv \frac{e}{\Omega}\sum_\mathbf{k} f_\mathbf{k} (\mathbf{r},t),\label{eq:rhofactorization}
   \\
    \vb{j}(\vb{r},t) &\equiv \frac{e}{\Omega}\sum_\mathbf{k} \vb{v}_{\vb{k}}f_\mathbf{k} (\mathbf{r},t)\label{eq:factorization}.
\end{align}
For the macroscopic lattice displacement $\vb{u}$, we use the definition in Eq.~\eqref{eq:displacement}. 
In the case of a large sphere in comparison to the size of a unit cell, we regard the lattice vectors $\mathbf{R}_n$ as continuous and will replace them with the continuous space variable $\vb{r}$, i.e., $\mathbf{R}_n\rightarrow \mathbf{r}$.\\
Combining Eq.~\eqref{eq:microscopic_oscillator} for the coherent phonon amplitude with the definition of the lattice displacement in Eq.~\eqref{eq:displacement} and the densities in Eqs.~(\ref{eq:rhofactorization},\ref{eq:factorization}), we obtain an equation of motion for the displacement,
\begin{align}
    \qty[\partial_t^2 + 2\gamma_{\text{ph}} \partial_t - c_{LA}^2\nabla^2_{\vb{r}}]
    &\vb{u}(\vb{r},t)=\label{eq:eomlatticedislong}\\ 
    \frac{V_0}{2M}
    \nabla_{\vb{r}} \sum_\mathbf{k} f_\mathbf{k} (\mathbf{r},t)&\nonumber
    +\frac{6h}{\sqrt{2MN\hbar}}\frac{\vb{r}}{\ell}
	\sum_{\vb{q}}
	\frac{1}{\omega_{\vb{q}}}n_{\vb{q}}^c(\vb{r},t).
\end{align} 
At this stage, we assumed that the microscopic scattering term $\gamma_{\vb{q}}$ can be approximated by a macroscopic, overall momentum-independent term $\gamma_{\text{ph}}$. 
Furthermore, we have assumed that the major momentum dependence of the matrix element $\Tilde{h}_{\vb{qq'q''}}$ is contained in the dispersion $\omega_{\vb{q}}$, such that it can be approximated as $\Tilde{h}_{\vb{qq'q''}} \approx h/\sqrt{\omega_{\vb{q}}\omega_{\vb{q'}}\omega_{\vb{q''}}}$\cite{gross1986many} with the momentum independent quantity $h$. 
In addition, we made use of the symmetry $\omega_{\vb{q}}=\omega_{\vb{-q}}$ of the dispersion relation.
The left-hand side accounts for the wave propagation of the phonon with the velocity of sound $c_{LA}$ and the phonon-phonon interaction induced damping term $\gamma_{\text{ph}}$.\cite{madelung1978introduction}
The right-hand side accounts for the sources of the coherent oscillations where we identify two different contributions. 
The first term accounts for the displacement of the lattice vectors via spatial gradients of the electron density as defined in Eq.~\eqref{eq:factorization}. 
The second term originates from the anharmonic phonon-phonon interaction where we incorporated the constant term in Eq.~\eqref{eq:microscopic_oscillator} as an offset of the oscillator position. 
Here, the parameter $\ell$ is the binding length of the material, which had to be included as the anharmonic thermal expansion in Eq.~\eqref{eq:eomlatticedislong} due to phonon-phonon interaction is an extensive quantity that increases with particle size compared to the harmonic contribution of the electron-phonon interaction in the first source term in Eq.~\eqref{eq:eomlatticedislong}.\\
In the oscillation Eq.~\eqref{eq:eomlatticedislong}, we find the thermal expansion to contribute in terms of the incoherent phonon distribution $n_{\vb{q}}^c(\vb{r})$. 
In the following, we will assume the distribution to be spatially homogeneous and to follow a Bose-Einstein distribution,
\begin{align}
	\frac{6h}{\sqrt{2MN\hbar}}\frac{\vb{r}}{\ell}
	\sum_{\vb{q}}
	\frac{1}{\omega_{\vb{q}}}n_{\vb{q}}^c(\vb{r})
	\approx\frac{18h\sqrt{N}}{\sqrt{2M\hbar^3}}\frac{\vb{r}}{\omega_D^2 \ell}k_B T. \label{eq:definitions}
\end{align}
The sum can be performed in Debye approximation by replacing the sum over the first Brillouin zone with an integral over a sphere with the radius $q_D$ which is defined such that the amount of states is equivalent to the amount of atoms. 
Using the linear dispersion of the Debye approximation, the Debye momentum $q_D$ can be replaced by the Debye frequency $\omega_D$.\cite{czycholl2017theoretische} 
We argue that the temperature of the nanoparticle can be approximated as spatially homogeneous as the particle is small compared to the optical wavelength, and the momentum-relaxation process is fast compared to the pulse width in the hydrodynamic limit.\\
As only temperature changes drive the nanoparticle oscillation and the initial temperature contribution in an oscillator equation is just an offset of the equilibrium position at the initial temperature $T_{\text{eq}}$ before optical excitation, in the oscillator equation only temperature differences $\Delta T(t) = T(t)-T_{\text{eq}}$ influence the dynamics,
\begin{align}
    &\qty[\partial_t^2 + 2\gamma_{\text{ph}} \partial_t - c_{LA}^2\nabla^2_{\vb{r}}]
    \vb{u}(\vb{r},t)=\label{eq:eomlatticedis}
    \beta 
    \nabla_{\vb{r}} \rho(\mathbf{r},t)
    +\boldsymbol{\xi}\Delta T(t),
\end{align} 
with the definitions, 
\begin{align}
    \beta \equiv \frac{V_0 \Omega}{2Me}, 
    \qquad
    \boldsymbol{\xi} \equiv \frac{18h\sqrt{N}}{\sqrt{2M\hbar^3}}\frac{\vb{r}}{\omega_D^2 \ell}k_B.
    \label{eq:prefactors}
\end{align}
We will return to these constants in App.~\ref{sec:approximations} where suitable approximations will be made to estimate their respective size.\\
In order to derive an equation of motion for the electron density, we sum Eq.~\eqref{eq:eomwignere} over momenta $\vb{k}$ in the spirit of the hydrodynamic approach in Eq.~\eqref{eq:rhofactorization}. 
As scattering terms conserve the local electronic density in the gradient approximation, they vanish under the momentum sum. This results in a continuity equation for the electron density $\rho(\vb{r},t)$ and the current density $\vb{j}(\vb{r},t)$,
\begin{align}
\partial_t \rho (\mathbf{r},t) + \nabla \cdot \mathbf{j} (\mathbf{r},t) = 0.\label{eq:conti}
\end{align}
The appearing current density $\vb{j}(\vb{r},t)$ is defined in Eq.~\eqref{eq:factorization}.
To obtain a closed set of equations, we derive an equation of motion for the current density from Eq.~\eqref{eq:eomwignere} by multiplying the microscopic dynamical equation \eqref{eq:eomwignere} with the velocity and summing over all momenta. 
We find a generalized Euler equation that also includes coherent phonon oscillations as source terms for the electron current density in the MNP,
\begin{align}
    \partial_t \mathbf{j} (\mathbf{r},t) 
        &= 
        - \gamma_v (T,\rho)\mathbf{j} (\mathbf{r},t)
    -\nabla \cdot \underline{\underline{\vb{P}}} (\mathbf{r},t)\label{eq:euler}
         \\\nonumber
     &-\rho (\mathbf{r},t) 
    \left( 
        \frac{e}{m}\Tilde{\vb{E}}(\vb{r},t)
        - \frac{V_0 N}
            {m} 
        \nabla \otimes \nabla \cdot \mathbf{u} (\mathbf{r},t) 
    \right).
\end{align}
The first term accounts for the decay of the macroscopic current density. 
Here, we have introduced the decay constant $\gamma_v (T,\rho)$, which has to be determined from the electron-phonon scattering contribution to Eq.~\eqref{eq:eomwignere} as the effective relaxation time of the current density.\cite{czycholl2017theoretische} 
It depends on the temperature as well as on the electron density itself.
For weak excitation as discussed here, only the temperature dependence is relevant.
The second term is the divergence of the second-order momentum in the factorization procedure of the Wigner function, which is the Cauchy stress tensor 
$\underline{\underline{\vb{P}}} \equiv \frac{e}{\Omega}\sum_\mathbf{q} \mathbf{v}_\mathbf{q}\otimes\mathbf{v}_\mathbf{q} f_\mathbf{q} (\mathbf{r},t)$. 
The second line accounts for the acceleration of the electrons in the total electric field $\Tilde{\vb{E}}$ or via the lattice displacement field $\vb{u}$, respectively.
In the next step, we decompose the velocity $\vb{v}_{\vb{q}} = \vb{v}+\delta\vb{v}_{\vb{q}}$ in a momentum-independent mean-field contribution and a correction. 
Accordingly, the Cauchy stress tensor in Eq.~\eqref{eq:euler} reveals two contributions. 
The first kinetic term is a tensor product of the mean-field velocity with itself, and the second one arises as a correction to the second-order momenta of the Wigner function which we identify as the pressure contribution 
$\underline{\underline{\hat{\vb{P}}}} \equiv \frac{e}{\Omega}\sum_\mathbf{q} \delta\mathbf{v}_\mathbf{q}\otimes\delta\mathbf{v}_\mathbf{q} f_\mathbf{q} (\mathbf{r},t)$. 
This way, the velocity is promoted to a velocity field $\vb{v}(\vb{r},t)$ which allows us to factorize the macroscopic current density $\mathbf{j} (\mathbf{r},t) = \rho (\mathbf{r},t)\, \mathbf{v} (\mathbf{r},t)$ into electron density $\rho (\mathbf{r},t)$ and velocity field $\mathbf{v} (\mathbf{r},t)$ of the electrons. 
Applying this decomposition to the continuity and Euler equation, we obtain, after some algebraic transformations, the final set of continuity equation,
\begin{align}
    \partial_t \rho(\vb{r},t)+\nabla\cdot \left(\rho (\mathbf{r},t) \vb{v} (\mathbf{r},t)\right)&=0\label{eq:conti2},
\end{align}
and Euler equation,
\begin{align}
        &\biggl[ 
            \partial_t 
            + \vb{v} (\mathbf{r},t)\cdot \nabla 
        \biggr] 
        \vb{v} (\mathbf{r},t)       
            =\label{eq:euler2}\\
       &-\gamma_v(T,\rho)\, \vb{v} (\mathbf{r},t)
      -\flatfrac{\nabla \cdot \underline{\underline{\hat{\vb{P}}}}(\vb{r},t)}{\rho(\vb{r},t)}
       \nonumber\\
       &-\frac{e}{m}\,\Tilde{\vb{E}}(\vb{r},t)
        +\frac{V_0 N}{m} 
                \nabla \otimes \nabla \cdot \mathbf{u} (\mathbf{r},t).\nonumber
\end{align}
The left-hand side in Eq.~\eqref{eq:euler2} accounts for the substantial derivative of the velocity field. 
The first term on the right-hand side describes the decay of the velocity due to interaction with the phonons. 
The second term accounts for the source of the velocity in the presence of pressure. 
The pressure tensor will be expressed in terms of the scalar pressure function $P(\vb{r},t)$ by $\underline{\underline{\hat{\vb{P}}}}(\vb{r},t) = P(\vb{r},t)\mathbbm{1}$\cite{shi2021ideal} in the following. 
The last line in Eq.~\eqref{eq:euler2} considers the acceleration of the electrons caused by the self-consistent near-field electric field $\Tilde{\vb{E}}$ and the lattice displacement $\vb{u}$. 
In the remaining calculation, this latter term will be neglected since we assume a dominant acceleration of the electrons by the electric field and the perturbation by $\vb{u}(\vb{r},t)$ to contribute only in higher order of the electric field $\Tilde{\vb{E}}(\vb{r},t)$.\\
All in all, Eqs.~(\ref{eq:eomlatticedis},\ref{eq:conti2},\ref{eq:euler2}) have to be solved self-consistently.
To approach Eq.~\eqref{eq:euler2} analytically, it will be coarse-grained in time.
In our system, the optical excitation leads to a non-equilibrium carrier distribution whose energy is thermalized by electron-electron scattering between electrons on very short time scales compared to the electron-phonon interaction.\cite{wilson2020parametric}  
On these short timescales, i.e., already during and shortly after the pulse, the density gradient $\nabla_{\vb{k}}f_{\vb{k}}$ builds up. 
The fast thermalization process allows us to find a local equilibrium description for the electrons on time scales of the electron-phonon coupling, which means that we can define a local equation of state with time-varying temperature $T(t)$ determined by the electron-phonon interaction.
Therefore, we assume an expression for the electron pressure in terms of the equation of state for fermions:
The pressure for a free Fermi gas can be given as\cite{bloch1933bremsvermoegen,raza2015nonlocal,moeferdt2018plasmonic}
\begin{align}
    P  (\vb{r},t)  = \kappa \rho^{5/3}  (\vb{r},t) \label{eq:press},
\end{align}
where the proportionality constant $\kappa = \frac{\hbar^2}{5me^{5/3}}\qty(3\pi^2)^{2/3}$ is adapted from the typical Fermi gas constant\cite{bloch1933bremsvermoegen} to our calculations for the charge density.
\section{Linearization of the electron dynamics}
\label{sec:linearization}
%
In order to solve the nonlinear set of equations (\ref{eq:conti2},\ref{eq:euler2}) and to use the results to determine the coherent phonon field $\vb{u}(\vb{r},t)$ in Eq.~\eqref{eq:eomlatticedis}, we expand the respective quantities in orders of the electric field,\cite{moeferdt2018plasmonic}
\begin{align}
    \rho(\vb{r},t) &= \rho_0 + \rho_1 (\vb{r},t) +\rho_2 (\vb{r},t) + \mathcal{O}(\vb{E}^3),\\
    \vb{v}(\vb{r},t) &=  \vb{v}_0 + \vb{v}_1 (\vb{r},t) + \vb{v}_2 (\vb{r},t)  + \mathcal{O}(\vb{E}^3).
\end{align}
For better readability, the spatial and temporal dependencies of the observables are suppressed from now on.
The density dependence of the pressure term in Eq.~\eqref{eq:press} will be approximated as\cite{moeferdt2018plasmonic}
\begin{align}
    \rho^{5/3} \approx \rho_0^{5/3} + \frac{5}{3} \rho_0^{2/3}\rho_1 + \frac{5}{9}\rho_0^{-1/3}\rho_1^2 + \frac{5}{3}\rho_0^{2/3}\rho_2.
\end{align}
In the same way, we expand Eqs.~(\ref{eq:conti2},\ref{eq:euler2}) in orders of the electric field. In zeroth order in the electric field, we obtain
\begin{align}
    \partial_t \rho_0 &= 0, \\
    \vb{v}_0  &= 0,
\end{align}
which means that in the absence of the electric field, the velocity vanishes and the electron distribution remains unchanged.
The first-order equation of motion for the electron density reads 
\begin{align}
    \partial_t \rho_1  + \rho_0\nabla \cdot \vb{v}_1 & = 0,
    \end{align}
which is analogous to the continuity equation [Eq.~\ref{eq:conti2}]. 
The equation of motion for the first-order velocity is given as
\begin{align}
	\rho_0 \left( \partial_t + \gamma_v \right) \vb{v}_1 & = -\frac{5\kappa}{3} \rho_0^{2/3}\nabla\rho_1 - \frac{e}{m}\rho_0 \Tilde{\vb{E}}.\label{eq:firstorder}
\end{align}
Since the velocity scales linearly with the total electric field, it oscillates with the optical frequency and is typically too fast to be detected directly.
Before we derive the second order contributions, i.e., the electron density oscillations, which determine the experimental signals, we introduce the geometrical constraints that renormalize the electric field inside the metal nanoparticle. 
This can be determined self-consistently and results in a resonance shift of the nanoparticle onto the plasmon resonance 
$\omega_{\text{pl}} = \flatfrac{\omega_{\text{p}}}{\sqrt{\varepsilon_{\text{b}} + 2\varepsilon_{\text{out}}}}$\cite{haug2009quantum} 
with the plasma frequency $\omega_{\text{p}} = \qty(\flatfrac{ne^2}{m\varepsilon_0})^{1/2}$, the permittivity of the surrounding medium $\varepsilon_{\text{out}}$, and the dielectric constant that accounts for the screening by bound charges $\varepsilon_{\text{b}}$ inside the MNP. 
A more detailed description is given in App.~\ref{sec:selfconsistentfield}. We find from Eq.~\eqref{eq:firstorder},
\begin{align}
	\rho_0 \left( \partial_t + \gamma_v  + i\omega_{\text{pl}}\right) \vb{v}_1 & = -\frac{5\kappa}{3} \rho_0^{2/3}\nabla\rho_1 - 3\varepsilon_0\varepsilon_{\text{out}}\omega_{\text{pl}}^2\vb{E}_0,\label{eq:firstorder_second}
\end{align}
where, in addition to the plasmon frequency $\omega_{\text{pl}}$, the renormalized total electric field $\Tilde{\vb{E}}$ was expressed in terms of the screened externally applied field: 
$\Tilde{\vb{E}} \rightarrow \flatfrac{3\varepsilon_{\text{out}}}{(\varepsilon_{\text{b}}+2\varepsilon_{\text{out}})}\vb{E}_0$, which is equivalent to the equation we derived in Eq.~\eqref{eq:firstorder_currentdensity} in App.~\ref{sec:selfconsistentfield}. 
When deriving second order quantities in the following, we apply the notation 
$\Tilde{\vb{E}} \equiv \flatfrac{3\varepsilon_{\text{out}}}{(\varepsilon_{\text{b}}+2\varepsilon_{\text{out}})}\vb{E}_0$.
The time evolution of the second-order electron density is given by
\begin{align}
	\partial_t\rho_2 &+ \rho_0\nabla\cdot \vb{v}_2  = -\nabla\cdot (\rho_1 \vb{v}_1),
\end{align}
and the equation of motion for the second-order velocity is 
\begin{align}
	\rho_0 \partial_t \vb{v}_2&+\gamma_v \rho_0 \vb{v}_2  +\frac{5\kappa}{3}\rho_0^{2/3}\nabla\rho_2  \label{eq:secondorder}\\\nonumber= &-\rho_1 \partial_t\vb{v}_1-\rho_0 (\vb{v}_1 \cdot \nabla)\vb{v}_1 -\frac{5\kappa}{9}\rho_0^{-1/3}\nabla\rho_1^2
	\\\nonumber&+\frac{e}{m}\rho_1 \Tilde{\vb{E}} - \gamma_v \rho_1 \vb{v}_1.
\end{align}
In the following, we assume the carrier frequency $\omega_{\text{opt}}$ of the total electric field $\Tilde{\vb{E}}$ to be in resonance with the plasmon frequency $\omega_{\text{pl}}$. 
To extract the signals relevant for the optical detection that are proportional to the cycle-averaged intensity, we separate the slowly varying components $\Tilde{\vb{E}}^\pm  (\vb{r},t)$ from the fast oscillation by
\begin{align}
    \vb{\Tilde{E}} (\vb{r},t) &= 
        \frac 12 \qty[\Tilde{\vb{E}}^+  (\vb{r},t) e^{i\omega_{\text{opt}} t}
        +\Tilde{\vb{E}}^-   (\vb{r},t) e^{-i\omega_{\text{opt}} t}],\\
    \rho_1 (\vb{r},t)&= 
        \frac 12 \qty[\Tilde{\rho}_1^+(\vb{r},t)e^{i\omega_{\text{opt}} t}
        +\Tilde{\rho_1}^-(\vb{r},t)e^{-i\omega_{\text{opt}} t}],\\
    \vb{v}_1(\vb{r},t) &= 
        \frac 12 \qty[\Tilde{\vb{v}}_1^+(\vb{r},t)e^{i\omega_{\text{opt}} t}
        +\Tilde{\vb{v}}_1^-(\vb{r},t)e^{-i\omega_{\text{opt}} t}].
\end{align}
This allows us to move to a rotating frame and separate the slowly varying quantities in Eq.~\eqref{eq:secondorder}. 
We obtain 
\begin{align}
	\biggl[\partial_t - \frac{5}{3}\frac{\kappa}{\gamma_v}&\rho_0^{2/3}\nabla^2\biggr]\Tilde{\rho}_2\\\nonumber
	= \frac{1}{2}\nabla&\cdot
	\biggl\{
	    \frac{\rho_0}{\gamma_v}\qty[(\Tilde{\vb{v}}_1^+\cdot \nabla)\Tilde{\vb{v}}_1^-
	    +(\Tilde{\vb{v}}_1^-\cdot \nabla)\Tilde{\vb{v}}_1^+]
	\\\nonumber
	&-\frac{e}{m\gamma_v}\qty(\Tilde{\rho}_1^+ \Tilde{\vb{E}}^-
	+\Tilde{\rho}_1^- \Tilde{\vb{E}}^+)\\\nonumber
	&+\frac{1}{\gamma_v}\Tilde{\rho}_1^+(\partial_t-i\omega_{\text{opt}})\Tilde{\vb{v}}_1^-
	+\frac{1}{\gamma_v}\Tilde{\rho}_1^-(\partial_t+i\omega_{\text{opt}})\Tilde{\vb{v}}_1^+\\\nonumber
	&+\frac{10}{9}\frac{\kappa}{\gamma_v}\rho_0^{-1/3}
	\qty(\Tilde{\rho}_1^+\nabla\Tilde{\rho}_1^-
	+\Tilde{\rho}_1^-\nabla \Tilde{\rho}_1^+)\biggr\}.
\end{align} 
We focus on the dominant driving contribution by making use of the relation $\gamma_v \ll \omega_0$. 
With the additional assumption that left- and right-handed quantities are equivalent in magnitude, i.e., $\Tilde{\vb{E}}^+=\Tilde{\vb{E}}^-=\hat{\vb{E}}$, we are left with
\begin{align}
    	\qty[\partial_t - D\nabla^2]\Tilde{\rho}_2
	= &
	K\,\nabla\cdot\qty[(\hat{\vb{E}}\cdot \nabla)\hat{\vb{E}}
	].
	\label{eq:eomdensity}
\end{align}
Hence, we have identified a diffusion equation for the second-order electron density distribution that is driven by spatial gradients of the electric field intensity similar to the ponderomotive force. 
The prefactors are defined as $D = \flatfrac{5\kappa\rho_0^{2/3}}{3\gamma_v}$ and $K = \flatfrac{e^2\rho_0}{\gamma_v^3 m^2}$.
\section{Lamb modes for MNP oscillations}
\label{sec:lamb}
%
In order to solve the two coupled partial differential equations \eqref{eq:eomlatticedis} and \eqref{eq:eomdensity}, we expand the equations into vibrational eigenmodes of the MNP. 
These free oscillations of a sphere were first described in Ref.~\onlinecite{lamb1881vibrations}, solving for the modes of the vibrational equation for the continuum field $\vb{u}(\vb{r},t)$,
\begin{align}
    (\lambda+\mu)\nabla (\nabla\cdot \vb{u})+ \mu\nabla^2 \vb{u} = \rho\partial_t^2 \vb{u}
    \label{eq:lambequation},
\end{align}
where $\mu$ and $\lambda$ are the Lamé constants that are related to the elasticity tensor. 
The equation can be solved using three so-called Helmholtz potentials that satisfy Helmholtz equations. 
As a first approach, we consider the most symmetric potential only and thus restrict our calculation to the first Helmholtz potential $\phi_{\text{nlm}}(\vb{r})$,\cite{lamb1881vibrations,love1892treatise} which allows us to express the displacement $\vb{u}(\vb{r})$ as
\begin{align}
    \vb{u}_{\text{nlm}} (\vb{r})&= \nabla \phi_{\text{nlm}}(\vb{r}),\\
    \text{with}\qquad
    \phi_{\text{nlm}}(\vb{r})&= j_l(k_n r)P_l^m(\cos{\theta}) \exp{im\varphi}\label{eq:helmholtzpotential}.
\end{align}
The radial dependence is given by the spherical Bessel functions $j_l(k_n r)$, where $k_n$ are the radial wave numbers that are determined by the boundary condition. 
Implementing a stress-free boundary condition, $k_n$ is the $n$th solution of the boundary condition [cf.~Eq.~\ref{eq:boundary}] that is given in terms of the unitless frequency $\eta$,\cite{lamb1881vibrations,hartland2002coherent} 
\begin{align}
	\eta \cot{\eta} = 1- \frac{\eta^2}{4\delta^2}\label{eq:boundary}.
\end{align}
$\delta$ is the ratio of the transverse and longitudinal speed of sound $c_{\text{TA}}/c_{\text{LA}}$. 
The frequency $\omega$ is connected to the unitless frequency $\eta$ by the relation $\omega_n = c_{LA} \flatfrac{\eta_n}{R}$, where $R$ is the radius of the sphere. 
To give an example for purely radial modes, we discuss the $l=0$ case. 
Introducing units similar to Eq.~\eqref{eq:displacement} in  the homogeneous equation \eqref{eq:lambequation}, we find
\begin{align}
	\phi_n(r) &= \qty(\frac{\hbar}{2M\omega_n})^{\frac{1}{2}} \frac{1}{k_n}j_0(k_n r),\\
	\vb{u}_n(r) &= \qty(\frac{\hbar}{2M\omega_n})^{\frac{1}{2}} \frac{1}{k_n}\partial_r\qty(j_0(k_n r))\vb{e}_r.
\end{align}
As the wavenumbers $k_n$ originate from the stress-free boundary condition imposed by Eq.~\eqref{eq:boundary}, the spherical Bessel functions are not orthogonal. 
As an approach to the complex system of partial differential equations Eqs.~\eqref{eq:eomlatticedis} and \eqref{eq:eomdensity}, we expand the displacement $\vb{u}(\vb{r},t)$ and the Helmholtz potential $\phi(\vb{r},t)$ in terms of the fundamental modes $\vb{u}_0(\vb{r})$ and $\phi(\vb{r},t)$ of the system related to the first root of the boundary equation \eqref{eq:boundary},
\begin{align}
    \rho(\vb{r},t) &= \rho(t)\,\phi_0(\vb{r}),\\
    \vb{u}(\vb{r},t) &= a(t)\,\vb{u}_0(\vb{r}).
\end{align}
This leads to
\begin{align}
    &\qty[\partial_t^2+2\gamma_{\text{ph}} \partial_t +\omega_{0}^2]a(t) = \beta \rho(t) +\xi \Delta T(t)\label{eq:displace},\\
    &\partial_t \rho(t)+ D \frac{\omega_0^2}{c_1^2} \rho(t) = -\frac{K}{B_0}\int \vb{u}_0^*(\vb{r}) \cdot\left[(\hat{\vb{E}}\cdot\nabla)\hat{\vb{E}}\right]\dd^3r.
	\label{eq:diffusioncoefficient2}
\end{align}
Here, we defined
\begin{align}
    \xi \equiv \frac{1}{A_0} \int \vb{u}_0^*\cdot \boldsymbol{\xi}\,\dd^3r
    \label{eq:scalar_xi}
\end{align}
and used the definitions of the overlap functions, 
\begin{align}
	A_0 \equiv  \int \vb{u}_0^*(\vb{r}) \cdot \vb{u}_0(\vb{r})\,\dd^3r,
	\qquad
	B_0 \equiv  \int \phi_0^*(\vb{r})\,\phi_0(\vb{r})\,\dd^3r,
\end{align}
where the integration is to be performed over the volume of the nanosphere.
The electronic driving term $\beta$ defined in Eq.~\eqref{eq:prefactors} and the thermal driving term $\xi$ [cf.~Eqs.~\eqref{eq:prefactors} and \eqref{eq:scalar_xi}] in the oscillator equation \eqref{eq:displace} depend on the electron-phonon and phonon-phonon interaction, respectively. 
The electronic driving term $\beta$ scales linearly with the screened Coulomb potential $V_0$ and the thermal driving factor $\xi$ is proportional to the anharmonic phonon-phonon potential $h$ [cf.~Eq.~\ref{eq:prefactors}]. 
Since both contributions are difficult to find in the literature, we use two approaches in App.~\ref{sec:thomasfermi} and App.~\ref{sec:thermalexpansion} to obtain suitable approximations. 
In addition, App.~\ref{sec:mietheory} provides a field description based on Mie theory and a model for the temperature dynamics that allow solving the set of coupled equations in App.~\ref{sec:twotemperature}. 
The equations describing the temperature dynamics can be found in Eqs.~(\ref{eq:electrontemperature}, \ref{eq:latticetemperature}). 
All used parameters are given in Tab.~\ref{table_parameters}.
\section{Numerical results}
\label{sec:numerics}
%
In this section, we address the solution of the two coupled ordinary differential equations of a damped oscillator for the MNP lattice vibration [Eq.~\ref{eq:displace}] and for the spatial mode coefficient of the electron density [Eq.~\ref{eq:diffusioncoefficient2}] driven by an external electric field. 
This set of equations generalizes previous models; \cite{hartland2002coherent,hodak2000photophysics} in particular, it exhibits a new source term for the onset of vibrational breathing modes. 
The left-hand side of Eq.~\eqref{eq:displace} contains two sources: 
a direct driving term that follows the excited electron dynamics $\rho(t)$ (ponderomotive force) and a term describing the known thermal expansion following the temperature dynamics $\Delta T(t)$. 
With femtosecond accuracy, the equation can be evaluated numerically using the Runge-Kutta method for various pump fluences and associated electric fields. 
The two inhomogeneities of Eq.~\eqref{eq:displace} describe (i) the optically induced electron density gradients and (ii) changes in lattice temperature. 
These two independent sources allow us to selectively study the individual or joint effect of the driving terms by artificially turning the driving terms in Eq.~\eqref{eq:displace} on or off.

\begin{figure}
    \centering
    \includegraphics[width=\linewidth]{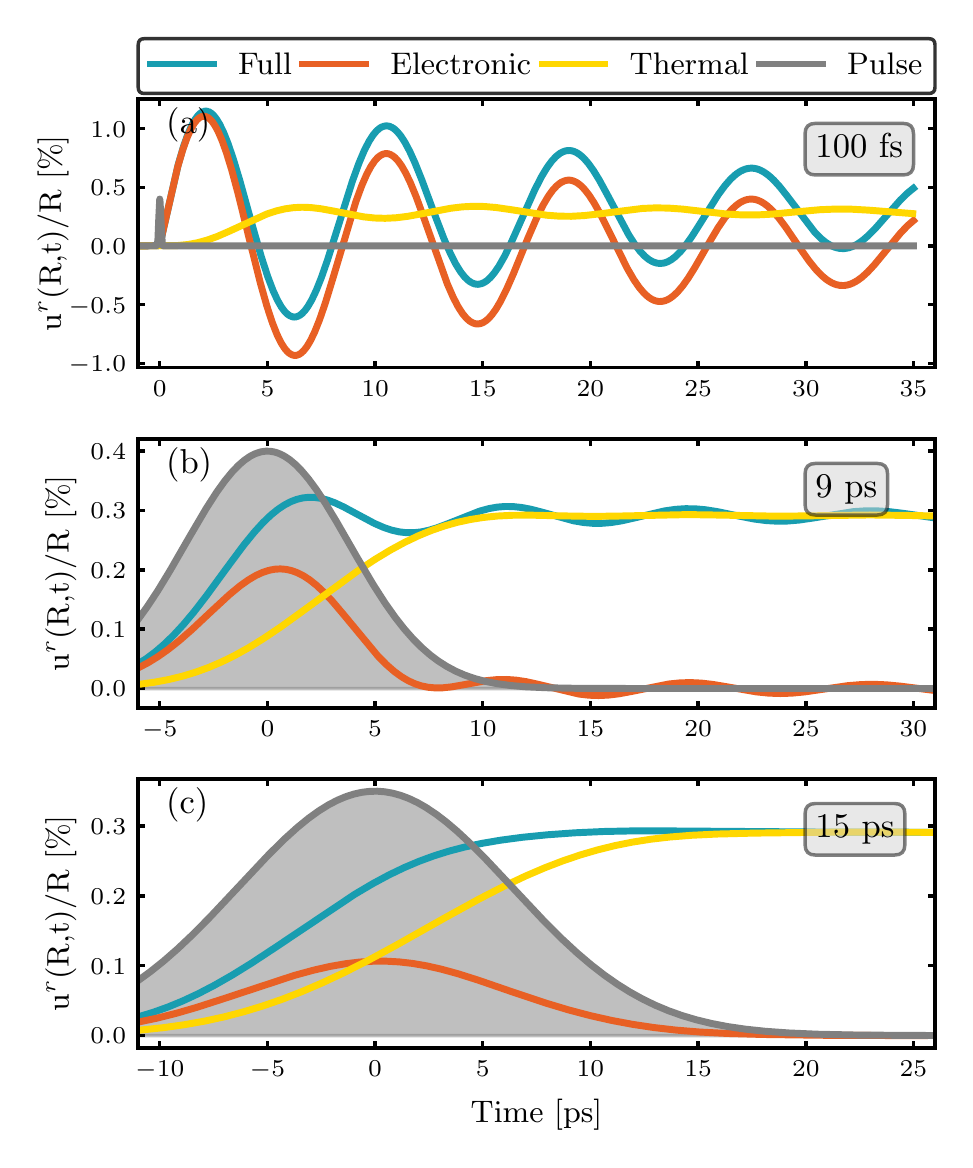}
    \caption{\textbf{Comparison of the impact of the individual coupling mechanisms on the amplitude and onset of the radial breathing mode oscillation.} In (a-c), we consider the thermal and electronic driving individually and combined ("full") for varying pulse widths that are given in the top right of the figures. For short pulses (a), the electronic driving term initiates the oscillation immediately with the optical pump whereas the pure thermal effect drives the oscillation with a delay of $\approx \unit[5]{ps}$ with a much smaller amplitude. Broader pulses with the same pump fluence cause smaller (b) or vanishing (c) oscillatory behavior. All pulses contain the same amount of energy and are displayed in arbitrary units.}
    \label{fig:displacement}
\end{figure}
Figs.~\ref{fig:displacement}(a-c) show the relative lattice displacement $u^r(\vb{R},t)/R$ for the individual influence of both, the electronic density gradient and thermal contribution and for the combined effect of both, where $u^r$ is the projection of the oscillation vector on the radial unit vector $u^r = \vb{e}_{r}^* \cdot \vb{u}$. 
From (a)-(c), we study the influence of the pulse width of the incident light field while keeping the pump fluence constant:
(a) pulse is much shorter than the period of the MNP size oscillation $\omega_{0}^{-1}$, (b) pulse length and period are similar, and (c) pulse length is longer than the oscillation period. 
This way, the pulse width distinguishes also between the different timescales involved by the different sources.\\ 
Looking at the onset of the oscillation in Fig.~\ref{fig:displacement}(a), we note that the electronic source term leads to an immediate increase in the lattice displacement by up to $\unit[1]{\%}$ within the first $\unit[5]{ps}$, while the thermally driven part hardly changes its magnitude during the same time and exhibits a delayed switch on of the oscillations with respect to the incident pulse. 
A purely thermal source term needs more time to start the oscillation so with purely thermal interaction, the first oscillation maximum is reached only after about $\unit[7]{ps}$. 
The combined effect of both driving terms leads to an ultrafast oscillation onset directly with the optical excitation and a shift of the equilibrium position of the oscillation with increasing lattice temperature.\\
In Fig.~\ref{fig:displacement}(b), the pulse width and period of the MNP oscillation are similar, and it can be seen that the thermal expansion reaches a similar magnitude as in Fig.~\ref{fig:displacement}(a). 
However, there are no thermally induced oscillations, and also electronically induced oscillations are small. 
Therefore, thermal expansion can be categorized as the dominant effect. 
In Fig.~\ref{fig:displacement}(c), the pulse is longer than a full oscillation period of the MNP. 
As a result, only a displacement of the equilibrium position of the nanoparticle can be observed, while an oscillatory behavior is no longer visible.\\
The comparison shows that a short pulse leads to a rapid onset of the MNP oscillations in Fig.~\ref{fig:displacement}(a), while a temporally broad pulse, as shown in Fig.~\ref{fig:displacement}(c), mainly causes a shift of the equilibrium position due to thermal expansion. 
This behavior can be explained by the interaction mechanism of electrons and coherent phonons that is mediated by the gradients of - on the one hand, the electric field in Eq.~\eqref{eq:eomdensity} and, on the other hand, the electron density in Eq.~\eqref{eq:eomlatticedis}. 
In general, temporally broader pulses lead to smaller gradients and thus to decreasing oscillation amplitudes for an interaction system that is driven by spatial gradients.\\
In the limiting case of temporally long pulses compared to the period length of the oscillating system, this can be thought of as the adiabatic limit, where a continuous shift of the equilibrium position occurs due to slow temperature changes. 
Whether an oscillation is excited depends primarily on the ratio of the period length of the oscillating system and the pulse width. 
Only relatively short pulses can excite the system to oscillate.\\
Compared to the literature, it should be emphasized that we show that short pulses compared to the period of the oscillation lead to an oscillation onset directly with the optical excitation with no significant influence of the thermal source. 
Hence, the source term caused by gradients in the electron density can be identified as the dominant driving source in the early phases of the oscillation.
The values used for the numerical implementation can be found in Tab.~\ref{table_parameters}.
\begin{table}
\centering
 \caption{Material parameters for Ag used in the numerical implementation }
 \begin{tabularx}{\linewidth}{Xcc}
 \hline\hline
    Parameter & Value & Reference \\
    \hline
    $m$                     & \unit[5.6856800]{fs$^2$eV nm$^{-2}$}                         & \\
    $M$                     & \unit[107.8682 $\times$ 10439.60413]{fs$^2$eV nm$^{-2}$}     & \onlinecite{meija2016atomic} \\
    $c_{LA}$                & \unit[$3.65\times 10^{-3}$]{nm fs$^{-1}$}                        & \onlinecite{lide2004crc}, \onlinecite{de2020understanding} \\
    $c_{TA}$                & \unit[$1.61\times 10^{-3}$]{nm fs$^{-1}$}                        & \onlinecite{lide2004crc}, \onlinecite{de2020understanding}  \\
    $c$                     & \unit[299.792458]{nm fs$^{-1}$}                                  &  \\
    $\gamma_{v}$           & \unit[0.0625]{fs$^{-1}$}                                       & \onlinecite{yang2015optical} \\
    $\gamma_{ph}$           & \unit[$1.25\times10^{-4}$]{fs$^{-1}$}                          & \onlinecite{arbouet2006optical},\onlinecite{del1999coherent},\onlinecite{voisin2000time} \\
    $Z^{\text{eff}}$        & 6.7555 & \onlinecite{clementi1967atomic} \\
    $\alpha_{\text{cl}}$    & \unit[$18.9\times 10^{-6}$]{K$^{-1}$}                          & \onlinecite{lide2004crc} \\
    \hline
    $g$                     & \unit[$3.1\times 10^{11}$]{W K$^{-1}$ mol$^{-1}$}         & \onlinecite{groeneveld1995femtosecond}\\
    $C_l$                   & \unit[$25.35$]{J K$^{-1}$ mol$^{-1}$}                     & \onlinecite{lide2004crc} \\
    $\zeta$                 & \unit[$6.74\times 10^{-4}$]{J\,K$^{-2}$ mol$^{-1}$}       & \onlinecite{kittel2018introduction},\onlinecite{ashcroft1976solid} \\
    $\tau_s$                & \unit[$4\times 10^{5}$]{fs}                               & \hyperlink{calc}{$^{\text{a}}$} \\
   \hline\hline
 \end{tabularx}\label{table_parameters}
\flushleft
 {\footnotesize{
\hypertarget{calc}{{$^{\text{a}}$}} Used parameters from similar noble metals in  Ref.~[\onlinecite{hartland2002coherent}] }}
\end{table}

\section{Conclusion}
\label{sec:conclusion}
In conclusion, we have found that the microscopic description of the electron-phonon coupling in the MNP allows the inclusion of a direct interaction between optically induced spatial electron gradients and coherent phonons to initiate radial MNP oscillations. 
The spatial gradients of the electron density are driven by the incident optical field and appear beyond the thermal contributions as an additional driving term in the oscillator equation for the lattice displacement. 
Numerical evaluation of the modified oscillator equation revealed that this direct interaction term causes an immediate onset of nanoparticle oscillation for optical excitation as long as the pulse width is short compared to the MNP oscillation period. 
In this case, the thermal contribution to the oscillation mainly shifts the equilibrium position of the oscillation and cannot explain the immediate onset of the oscillations for sufficiently short pulses. 
Therefore, the dominant source for the onset of the radial oscillations on short time scales is the direct coupling between electrons and coherent phonons. 
The indirect interaction via the incoherent phonons does not play a dominant role in the early stages of the oscillation.
%
\section*{Acknowledgments}
%
We acknowledge fruitful discussions with Jonas Grumm, Dominik Christiansen, Lara Greten, Manuel Katzer, Joris Sturm (TU Berlin) and Yannic Staechelin (Uni Hamburg).\\
This work is supported financially by the Deutsche Forschungsgemeinschaft (DFG) through Project SE 3098/1-1 (Project No. 432266622) and the Cluster of Excellence 'CUI: Advanced Imaging of Matter' - EXC 2056 - project ID 390715994.
%
\section*{Author Declarations}
%
The authors have no conflicts to disclose.\\
 This article may be downloaded for personal use only. Any other use requires prior permission of the author and AIP Publishing. This article appeared in R. Salzwedel et al., J. Chem. Phys., \textbf{158} (6), 064107 (2023) and may be found at \url{https://doi.org/10.1063/5.0139629}.

\appendix
\section{Coherent phonon damping rates}
\label{sec:coherent}
%
The microscopic damping rate of the coherent phonon amplitude in Eq.~\eqref{eq:microscopic_oscillator} is given by
\begin{align}
    \label{eq:phonon_damping}
    \gamma_{\vb{q}} \equiv \frac{2\pi}{\hbar^2}
    \sum_{\vb{q'}}
        \abs{\hat{h}_{\vb{-q,q',q-q'}}}^2 
        &\qty[
            1+\Tilde{n}_{\vb{q'}}^c + \Tilde{n}_{\vb{q-q'}}^c
        ]\\\nonumber
        &\times\delta(\omega_{\vb{q'}}-\omega_{\vb{q-q'}}-\omega_{\vb{q}}),    
\end{align}
which results from phonon-phonon interaction. 
The phonon terms are actual occupations, and off-diagonal terms are neglected as higher-order contributions at this stage.

\section{Hartree-Fock contributions}
\label{sec:hartree_fock}
%
On the Hartree-Fock level, the Hamiltonian can also be written as an effective single-particle Hamiltonian. 
In first-order gradient expansion, this results in the following equation for the Wigner distribution:\cite{hess1996maxwell}
\begin{align}
    \label{eq:hartreefock_bloch}
    \partial_t f_{\vb{k}}(\vb{r}) &= 
    \frac{1}{\hbar}
    \qty[
    -\pdv{\mathcal{E}_{\vb{k}}(\vb{r})}{\vb{k}}
    \pdv{f_{\vb{k}}(\vb{r})}{\vb{r}}
    +\pdv{\mathcal{E}_{\vb{k}}(\vb{r})}{\vb{r}}
    \pdv{f_{\vb{k}}(\vb{r})}{\vb{k}}
    ],
\end{align}
where the spatiotemporal effective single particle energy is defined as
\begin{align}
    \label{eq:effective_energy}
    \mathcal{E}_{\vb{k}}(\vb{r}) \equiv \varepsilon_{\vb{k}} - \Phi^H(\vb{r}) + \Phi_{\vb{k}}^F(\vb{r}),
\end{align}
with the additional Hartree potential $\Phi^H(\vb{r})$ and the Fock potential $\Phi_{\vb{k}}^F(\vb{r})$. 
The Hartree potential reads
\begin{align}
    \label{eq:hartree_potential}
    \Phi^H(\vb{r}) \equiv -\frac{1}{e\,\Omega} \int \mathrm{d}^3\vb{r'}~ V({\vb{r-r'}}) \sum_{\vb{k'}} f_{\vb{k'}}(\vb{r'}) ~,
\end{align}
where $V(\vb{r-r'})$ is the Coulomb potential in real space for the respective geometry - in this case, a spherical solution.\cite{de2010optical,jackson1999classical} 
The Fock potential is given by
\begin{align}
    \label{eq:fock_potential}
    \Phi_{\vb{k}}^F(\vb{r}) \equiv -\frac{1}{e}\sum_{\vb{k'}\neq \vb{k}}V_{\vb{k'-k}}f_{\vb{k'}}(\vb{r}),
\end{align}
where $V_{\vb{k'-k}}$ is the momentum space Coulomb potential that is also found in the Hamiltonian in Eq.~\eqref{eq:hamiltonian}. 
Use of Eq.~\eqref{eq:effective_energy} in Eq.~\eqref{eq:hartreefock_bloch} leads to the additional terms found in Eq.~\eqref{eq:eomwignere}.
\section{Self-consistent electric field}
\label{sec:selfconsistentfield}

In order to treat the self-consistent electric field $\Tilde{\vb{E}}$ inside the sphere, we use Mie theory to incorporate the geometric boundary conditions, resulting from the dielectric environment. 
We distinguish between the externally applied field $\vb{E}_0$, the background polarization of the sphere $\vb{P}_b$, and the Drude polarization $\vb{P}_d$ of the conduction electrons inside the sphere. 
In the following, we denote the dielectric constants of the inner and outer environments of the MNP by $\varepsilon_{\text{b}}$ and $\varepsilon_{\text{out}}$. 
The self-consistent field within the nanoparticle then reads\cite{griffiths1999introduction}
\begin{align}
    \Tilde{\vb{E}} = \vb{E}_{0} - \frac{1}{3 \varepsilon_0 \varepsilon_{\text{out}}} \vb{P}_b - \frac{1}{3 \varepsilon_0 \varepsilon_{\text{out}}} \vb{P}_d \label{eq:etot_pb_pd}.
\end{align}
The background polarization density $\vb{P}_b$ is given by the total electric field in the sphere multiplied by the effective background susceptibility 
$\Tilde{\chi}_b = \chi_{\text{b}}-\chi_{\text{out}} = \varepsilon_{\text{b}} - \varepsilon_{\text{out}}$. $\Tilde{\chi}_{\text{b}}$ 
contains the correction of the susceptibility $\varepsilon_{\text{b}}$ inside the sphere compared to the surrounding $\varepsilon_{\text{out}}$,
\begin{align}
    \vb{P}_b = \varepsilon_0 \Tilde{\chi}_{\text{b}}  \Tilde{\vb{E}}.
\end{align}
With this, the total field can be written as
\begin{align}
    \Tilde{\vb{E}} &= \frac{3\varepsilon_{\text{out}}}{2\varepsilon_{\text{out}} +\varepsilon_{\text{b}}} 
    \qty[
        \vb{E}_{0} - \frac{1}{3 \varepsilon_0 \varepsilon_{\text{out}}} \vb{P}_d \label{eq:E_tot}
        ].
\end{align}
The relationship between Drude polarization density $\vb{P}_d$ and the self-consistent electric field $\Tilde{\vb{E}}$ is classically derived in Drude theory [cf.~Eq.~\eqref{eq:drude}] and is given by
\begin{align}
    \vb{P}_d =  -\frac{\varepsilon_0\,\omega_p^2}{\omega^2+\mathrm{i}\gamma_{v} \omega} \Tilde{\vb{E}},
\end{align}
with the plasma frequency $\omega_p = \sqrt{\flatfrac{ne^2}{m\,\varepsilon_0}}$. 
Inserting this definition in Eq.~\eqref{eq:E_tot}, an oscillator equation for the Drude polarization density in frequency space depending on the external electric field $\vb{E}_0$ is obtained,
\begin{align}
    (\omega^2+ \mathrm{i}\gamma_v\omega -\omega_{\text{pl}}^2) \vb{P}_d &= -3 \varepsilon_0 \varepsilon_{\text{out}} \omega_{\text{pl}}^2 \vb{E}_{0},
    \label{eq:polarization_oscillator}
\end{align}
where we identified the plasmon frequency $\omega_{\text{pl}}$,
\begin{align}
    \omega_{\text{pl}} =\frac{\omega_p}{\sqrt{\varepsilon_b+2\varepsilon_{\text{out}}}} ~.
\end{align}
and the fact that the external electric field is renormalized 
$\vb{E}_0 \rightarrow \frac{3\varepsilon_{\text{out}}}{\varepsilon_{\text{b}}+2\varepsilon_{\text{out}}}\vb{E}_0$.
Rewriting the polarization density in Eq.~\eqref{eq:polarization_oscillator} into an equation for the current density $\vb{j}$, with $\vb{j} \equiv \partial_t \vb{P}$, we find in rotating wave approximation ($\omega_{\text{pl}}\approx \omega$),
\begin{align}
    \partial_t \vb{j} = -\gamma_v \vb{j} - i\omega_{\text{pl}}\vb{j} - \frac{3\varepsilon_{\text{out}}}{\varepsilon_b+2\varepsilon_{\text{out}}}\frac{e\rho}{m}\vb{E}_0
    \label{eq:firstorder_currentdensity}
\end{align}
Eq.~\eqref{eq:firstorder_currentdensity} shows that the geometry of our structure renormalizes the resonance frequency of our system to the plasmon frequency $\omega_{\text{pl}}$. 
This is usually derived from the resonance of the polarizability $\alpha(\omega)$ as can be found in Eq.~\eqref{eq:polarizability}.\cite{jackson1999classical,kreibig2013optical,maier2007plasmonics} 

\section{Scattering Rates}
\label{sec:scatteringrates}

The electron-phonon scattering rates in Eq.~\eqref{eq:eomwignere} are given by\cite{hess1996maxwell,jago2019spatio,selig2020suppression,venanzi2021terahertz}
\begin{align}
    \Gamma^{\text{in}}_{\mathbf{k+q,k}} (\mathbf{r},t) &=\frac{2\pi}{\hbar^2} \sum_{\pm} |g_\mathbf{q}|^2 f_\mathbf{k+q} (\mathbf{r},t) \times  \\\nonumber
    &\times\left(\frac{1}{2}\pm \frac{1}{2} + n_\mathbf{q} (\mathbf{r},t) \right) \delta \left(\varepsilon_\mathbf{k} - \varepsilon_{\mathbf{k+q}}\pm \hbar\omega_\mathbf{q}\right),\\
    \Gamma^{\text{out}}_{\mathbf{k,k+q}} (\mathbf{r},t) &= \frac{2\pi}{\hbar^2}\sum_{\pm} |g_\mathbf{q}|^2 \left(1 - f_\mathbf{k+q} (\mathbf{r},t)\right) \times  \\\nonumber
    &\times\left(\frac{1}{2}\pm \frac{1}{2} + n_\mathbf{q} (\mathbf{r},t) \right) \delta \left(\varepsilon_\mathbf{k} - \varepsilon_{\mathbf{k+q}}\mp \hbar\omega_\mathbf{q}\right),
\end{align}
and the electron-electron scattering rates in Eq.~\eqref{eq:eomwignere} are given by\cite{hess1996maxwell,lindberg1988effective,binder1992carrier,selig2020suppression,venanzi2021terahertz}
\begin{align}
    \mathcal{W}^{\text{in}}_\mathbf{k+q,k}(\mathbf{r},t) &= \frac{2\pi }{\hbar^2}\sum_{\vb{k'}} \abs{V_{\vb{q}}}^2 f_{\vb{k+q}}(\vb{r},t) f_{\vb{k'}}(\vb{r},t)
    \times\\&\times
    \qty(1-f_{\vb{k'+q}}(\vb{r},t))
    \delta \left(\varepsilon_\mathbf{k+q} +\varepsilon_{\vb{k'}}- \varepsilon_{\mathbf{k'+q}}-\varepsilon_{\vb{k}}\right),\nonumber\\
    \mathcal{W}^{\text{out}}_\mathbf{k,k+q}(\mathbf{r},t) &= \frac{2\pi }{\hbar^2}\sum_{\vb{k'}} \abs{V_{\vb{q}}}^2 f_{\vb{k'+q}}(\vb{r},t)\qty(1-f_{\vb{k'}}(\vb{r},t))
    \times\\&\times
    \qty(1-f_{\vb{k+q}}(\vb{r},t))
    \delta \left(\varepsilon_\mathbf{k+q} +\varepsilon_{\vb{k'}}- \varepsilon_{\mathbf{k'+q}}-\varepsilon_{\vb{k}}\right).\nonumber
\end{align}
All processes depend on the occupation of the initial state of the corresponding electronic transition, i.e., on the electron occupation terms $f_{\vb{k}}$ in Eq.~\eqref{eq:eomwignere} and on the hole occupation terms $(1-f_{\vb{k}})$. 
The Dirac $\delta$ distributions that occur take into account the conservation of energy and momentum in electron-phonon or electron-electron scattering and in phonon emission and absorption processes.
For electron-phonon in- and out-scattering, we find contributions from the phonon emission ($+$) term in the summation and phonon absorption ($-$).
\section{Parameter estimation}
\label{sec:approximations}
%
In this section, we are concerned with the driving terms of the set of equations (\ref{eq:displace}, \ref{eq:diffusioncoefficient2}). 
In the following we will provide approximations or classical descriptions for the individual driving terms that allow estimating their influence. 
In Sec.~\ref{sec:thomasfermi}, the electronic driving term is approximated using a Thomas-Fermi theory for the screening. 
In Sec.~\ref{sec:thermalexpansion}, the thermal driving term is evaluated using a comparison to classical thermal expansion.  
Also, to evaluate Eqs.~\eqref{eq:eomlatticedis} and \eqref{eq:eomdensity}, we need to include the field description (Sec.~\ref{sec:mietheory}) based on Mie theory and calculate the overlap with the fundamental vibrational mode and the temperature dynamics, which will be done in Sec.~\ref{sec:twotemperature} with the use of the classical two temperature model.

\subsection{Thomas-Fermi theory for strong screening}
\label{sec:thomasfermi}

In deriving the oscillator equation in Sec.~\ref{sec:factorization}, we assumed that the electron-phonon potential is strongly screened, which is a good approximation for most metals.\cite{czycholl2017theoretische,madelung1978introduction} 
This allows us to use an effective electron-phonon coupling that is $\vb{q}$ independent. 
It reads\cite{kittel1987quantum}
\begin{align}
    V_0 = \frac{Z^{\text{eff}}e^2}{\varepsilon_0 \Omega k_s^2},
\end{align}
with an effective inverse screening length $k_s$ and an effective nuclear charge $Z^{\text{eff}}$. 
For $Z^{\text{eff}}$, the value for the $5s$ band \cite{clementi1967atomic} is used since it is the only band that is only partially filled in silver. 
In the subsequent calculation, this inverse screening length is approximated by the Thomas-Fermi wave vector $k_{\text{TF}}$, which is given as 
\begin{align}
    k_{\text{TF}}^2 = \frac{e^2m}{\varepsilon_0 \hbar^2 \pi^2}(3\pi^2 n)^{1/3},
\end{align}

with the electron density $n$.  Combining this with the definition of the potential, one can simplify the driving term $\beta$ in Eq.~\eqref{eq:prefactors} of the breathing oscillation due to electron gradients to
\begin{align}
    \beta = \frac{Z^{\text{eff}}\hbar^2 \pi^2}{2mMe}\qty(3\pi^2 n)^{-1/3},
    \label{eq:beta}
\end{align}
which depends only on the intrinsic quantities electron density $n$, effective electron mass $m$, and unit cell mass $M$ for the case of strong screening.
%
\subsection{Thermal expansion}
\label{sec:thermalexpansion}

Since the contribution of the anharmonic potential to the Hamiltonian [Eq.~\ref{eq:hamiltonian}] describes the macroscopic process of thermal expansion, we will approximate the corresponding prefactor in Eq.~\eqref{eq:displace} by the classical linear expansion coefficient $\alpha_{\text{cl}}$. 
This allows us to compare our oscillator equation with the classical description.\cite{del1999coherent,hartland2002coherent}
The linear expansion coefficients is given by 
\begin{align}
    \frac{u_r(\vb{R},t)}{\abs{\vb{R}}}
    =\alpha_{\text{cl}} \Delta T,
    \label{eq:linearexpansion}
\end{align}
with the fraction of the radial projection of the lattice displacement $\vb{u}(\vb{R},t)$ to the total particle radius $R$. 
Our calculated quantity $\vb{u}(\vb{R},t)$ is a measure for the absolute radial displacement at a certain lattice position. 
The position $\vb{R}$ is chosen to be on the surface of the nanoparticle.\\
Comparison of the thermal driving term in Eq.~\eqref{eq:displace} with the classical equivalent in Eq.~\eqref{eq:linearexpansion} reveals that by expanding in the fundamental Lamb mode $\vb{u}(\vb{r},t)=\vb{u}(\vb{r})a(t)$, we find
\begin{align}
    a(t)= \frac{\alpha_{\text{cl}} R \Delta T(t)}{A_0}\int \dd^3 r \,u_r^*(\vb{r}),
\end{align}
so that the oscillation Eq.~\eqref{eq:displace} can be expressed using the linear expansion coefficient $\alpha_{\text{cl}}$,
\begin{align}
    \qty[\partial_t^2+2\gamma_{\text{ph}} \partial_t +\omega_0^2]a(t) = \beta \rho(t) +\xi \Delta T(t)\label{eq:oscillator}, 
\end{align}
where the scalar coupling constant in Eq.~\eqref{eq:scalar_xi} can now be approximated as 
\begin{align}
    \xi \equiv \alpha_{\text{cl}} \frac{\omega_0^2 R \int\dd^3 r\, u_r^*(\vb{r})}{A_0}.
    \label{eq:xi}
\end{align}
This calculation leads to a similar thermal drive term as given in the oscillation equation in Ref.~\onlinecite{hartland2002coherent}. 
Therefore, it is expected that the models agree for purely thermal drive and lead to a similar oscillation behavior.
%
\subsection{Mie theory for electric field}
\label{sec:mietheory}

\begin{figure}[h!]
    \centering
    \includegraphics[width=\linewidth]{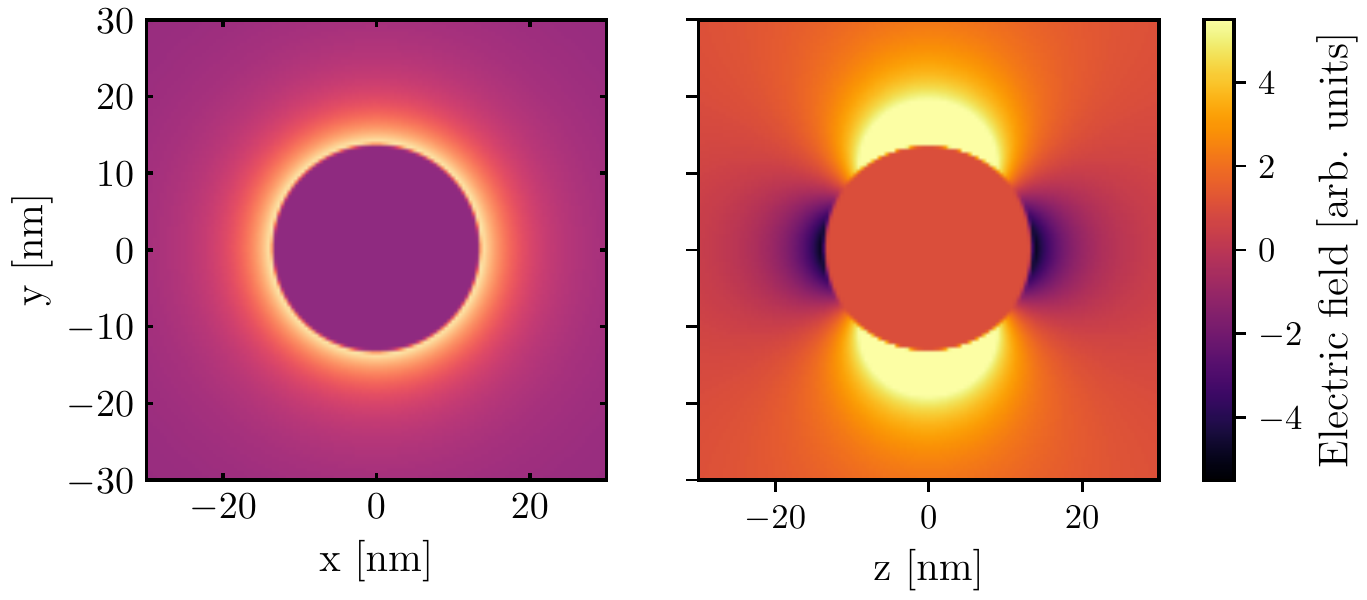}
    \caption{\textbf{Electric field of a spherical silver nanoparticle.} For a z-aligned incident electric field, the resulting z component is illustrated on the left for the xy-plane. To the right, we see the x component illustrated for the xz plane.}
    \label{fig:dipolefield}
\end{figure}

In the density equation \eqref{eq:diffusioncoefficient2}, we need to find a way to express the inhomogeneity in terms of the electric field. 
We use Mie theory in order to provide a good approximation for the field as the particle is much smaller in size compared to the wavelength of the incident light such that a quasistatic approximation is applicable. 
We use a Heaviside Theta function to unite the standard Mie field terms \cite{mie1908beitrage,jackson1999classical} in a single expression,
\begin{align}
	\vb{E}(\vb{r}) = \Theta(r-R)\,\vb{E}_{\text{out}} + \Theta(R-r)\,\vb{E}_{\text{in}},
\end{align}
with the individual definitions,
\begin{align}
	\vb{E}_{\text{in}} &= \frac{3\varepsilon_{\text{out}}}{\varepsilon(\omega) + 2\varepsilon_{\text{out}}}\vb{E}_0,\label{eq:electric_field_inside}\\
	\vb{E}_{\text{out}} &= \vb{E}_0 + \frac{3\vb{n}(\vb{n}\cdot\vb{p})-\vb{p}}{4\pi \varepsilon_0\varepsilon_{\text{out}} r^3}.
\end{align}
Some components of the electric field are illustrated in Fig.~\ref{fig:dipolefield}.

The dipole moment $\vb{p}$ is given by $\vb{p} = \alpha(\omega) \vb{E}$ where $\alpha(\omega)$ is the polarizability of the MNP,\cite{mie1908beitrage}
\begin{align}
    \label{eq:polarizability}
    \alpha(\omega) = 4\pi \varepsilon_0 \varepsilon_{\text{out}} R^3\frac{\varepsilon(\omega)-\varepsilon_{\text{out}}}{ \varepsilon(\omega)+2\varepsilon_{\text{out}}}.
\end{align}
where the frequency-dependent material permittivity is given by the Drude model,\cite{bohren2008absorption}
\begin{align}
    \varepsilon(\omega) = \varepsilon_{\text{b}} - \frac{\omega_p^2}{\omega(\omega+i\gamma_{v})},
    \label{eq:drude}
\end{align}
with the plasma frequency $\omega_p^2 = \flatfrac{n e^2}{\varepsilon_0 m}$, the electron dephasing $\gamma_{v}$, and the high-frequency value of the permittivity $\varepsilon_{\text{b}}$.\cite{maier2007plasmonics,yang2015optical}\\
Hence, the driving term in the density equation can be calculated to be 
\begin{align}
    -\frac{\rho_0 e^2}{\gamma_v^3 m^2 B_0}&\int \vb{u}_0^*(\vb{r}) \cdot\left[(\hat{{\vb{E}}}\cdot\nabla)\hat{\vb{E}}\right]\dd^3r\label{eq:electricfield}\\\nonumber
    &=
	-\frac{4\pi}{3} \qty(\frac{\hbar}{2m\omega})^{\flatfrac{1}{2}} \frac{\rho_0 e^2}{\gamma_v^3 m^2 B_0} \hat{E}^2 \times\\\nonumber
	&\times
	\biggl[
	\frac{\partial_r j_0(k_0 R)}{k_0}
	\biggl(
	(1-b^2)R^2 \\\nonumber
        &\qquad+ \frac{4\alpha(\omega_{\text{opt}}) \Tilde{k}}{R}+\frac{4\alpha^2(\omega_{\text{opt}}) \Tilde{k}^2}{R^4}
	\biggr)
	\biggr],
\end{align}
with the definitions $\Tilde{k} = \flatfrac{1}{4\pi \varepsilon_0\varepsilon_{\text{out}}}$ and $b = \flatfrac{3\varepsilon_{\text{b}}}{(\varepsilon(\omega_{\text{opt}})+\varepsilon_{\text{out}})}$. 
This provides the driving term for the electron gradient in the metal nanoparticle. As Fig.~\ref{fig:dipolefield} indicates, the main source of the electron density inhomogeneity originates from the boundary where the electric field changes most.
%
\subsection{Two-temperature model}
\label{sec:twotemperature}
%
The anharmonic phonon-phonon Hamiltonian [Eq.~\ref{eq:hamiltonian}] includes thermal effects in the coupled dynamics. 
In the oscillator equation \eqref{eq:displace}, changes in lattice temperature influence the equilibrium position of the ions and can cause transient oscillations for rapidly changing shifts in temperature. 
For our calculations, the temperature evolution will be modeled using the classical two-temperature model,\cite{kaganov1957relaxation,del1999coherent,hartland2002coherent}
\begin{align}
    C_e(T_e)
    \pdv{T_e}{t} 
    &= 
    -g(T_e - T_l) 
    + \frac{W_0}{\sqrt{\pi}\sigma} 
    \exp{-t^2/\sigma^2}\label{eq:electrontemperature},
    \\
    C_l
    \pdv{T_l}{t}
    &=
    g(T_e-T_l)
    -(T_l-298)/\tau_s\label{eq:latticetemperature},
\end{align}
which models the thermal energy transfer between the electrons and lattice via the coupling constant $g$ that is linked to quantum mechanical electron-phonon coupling\cite{allen1987theory} and can be accessed experimentally.\cite{staechelin2021size}
$T_e$ and $T_l$ are electron and lattice temperatures.
The optical excitation is described by the absorbed energy $W_0$, which can be determined from the absorbance measured in optical experiments, the pump fluence of the incident electric field, and the concentration of Ag nanoparticles. 
The heat capacities for the electrons $C_e(T_e) = \zeta\,T_e$ and the lattice $C_l$ as well as the time scale $\tau_s$ for the energy transfer to the surrounding can be found in Tab.~\ref{table_parameters}.

\bibliography{references}
\end{document}